# Thermodynamic stability of twisted domains in $AgCrSe_2$ thin films grown on lattice-matched YSZ(111) substrate

Haruto Sato[1], Kota Mihara[1], Kenshin Inamura[1], Yusuke Tajima[1],

Kazutaka Kudo[1,2], Jobu Matsuno[1,2], and Junichi Shiogai[1,2,†]

[1]*Department of Physics, The University of Osaka, Toyonaka, Osaka, 560-0043, Japan.*

[2]*Division of Spintronics Research Network, Institute for Open and Transdisciplinary Research Initiatives, The University of Osaka, Suita, Osaka, 565-0871, Japan.*

[†]The author to whom correspondence should be addressed.

junichi.shiogai.sci@osaka-u.ac.jp

**Abstract**

Control of structural domains in epitaxial thin films of functional materials is a fundamental technique to utilize their intrinsic physical and chemical properties in solid-state devices. In this study, we report on suppression of twisted-domain formation in thin-film growth of polar magnetic semiconductor $AgCrSe_2$ using pulsed-laser deposition. In exploring concomitant optimized growth temperature and Ag/Cr composition ratio of supply, we find the critical growth temperature ($T_{sub}$) for obtaining single 60° domain in *c*-axis oriented $AgCrSe_2$ thin film on a lattice-matched (111) plane of the yttria-stabilized zirconia substrate. At temperatures below and above the critical $T_{sub}$, metastable 0° domain in addition to the 60° domain emerges, indicating delicate energy balance of thermodynamic stability for obtaining the single-domain structure. Surface structural analysis using time-of-flight low-energy atom scattering spectroscopy reveals the presence of two polar orientations along $+Z$ and $-Z$ directions. These findings provide valuable insights into the thin-film growth mechanisms for a family of two-dimensional compounds with rhombohedral lattices.

**Main**

## I. INTRODUCTION

Control of structural domains in epitaxial thin films of quantum materials is a key technique for integrating their functionalities into electronic and photonic solid-state devices [1-10]. In general, the substrate materials available for thin-film growth have different crystal symmetries and lattice constants with epilayer materials. For example, the (0001) plane of $Al_2O_3$ and the (111) plane of cubic materials are often used for thin film growth of hexagonal and rhombohedral materials [7,8,11,12]. In such heteroepitaxy, growth mode of the first few atomic layers is a matter of interest because the subsequent precursors experience the underlying substance as homoepitaxy. Therefore, the environment of the substrate surface is a determining factor of the thermodynamic stability in physical and chemical absorptions of the precursors, and propagation of nucleated domains.

In the case of hexagonal and rhombohedral crystal structures, rotational degrees of freedom result in in-plane twisting of the epilayer domain [1,10,11]. In the case of the polar semiconductor ZnO (wurtzite, $P6_3mc$) deposited on the (0001) plane of $Al_2O_3$ (corundum, $R\bar{3}m$), it can have either aligned or 30°-twisted domains stabilized by the rotational commensurate epitaxy, depending on the substrate temperature and growth rate [1]. In addition, the in-plane twisting of ZnO is accompanied by the switching of the polar orientation, since the polarity is governed by the bonding configuration of the first epilayer, which is determined by rotational commensurate epitaxy. In the case of rhombohedral lattices, suppression of such twisted domains is even more difficult because the lattice mismatch is almost degenerate between 0° and 60° (equivalent to 180°) domains on threefold symmetric surface of the substrates [9,13]. Thus, formation of the single domain is realized by lowering threefold symmetry through the surface control of the substrate as exemplified for topological insulators $Bi_2Se_3$ (Refs. [4,5]) or two-dimensional semiconductor $MoS_2$ (Ref. [14]).

$AgCrSe_2$, a polar magnetic semiconductor, exhibits emergent spintronic functionalities such as a large magnetoresistance [15] and a spin-split band induced by the Rashba-type spin-orbit interaction due to spontaneous polarization [16]. $AgCrSe_2$ is also attractive as an energy harvesting material for its ultralow thermal conductivity [17] and superionic conduction [18,19]. These unique characters of $AgCrSe_2$ stem from the two types of the Ag atomic-layer structure intercalated between two-dimensional (2D) networks of edge-sharing $CrSe_6$ octahedra (denoted as $CrSe_2$ networks) as schematically shown in Fig. 1(a) and 1(b). It undergoes the second order transition from the low-temperature (low-$T$) phase ($R3m$) into the high-temperature (high-$T$) phase ($R\bar{3}m$) at the structural transition temperature $T_s$ between 450 K and 475 K. The low-$T$ phase lacks the inversion symmetry and polar orientation depends on the Ag site occupancy at either $\alpha$ (+Z polar) or $\beta$ (−Z polar) sites between the $CrSe_2$ networks as shown in Fig. 1(c). The high-$T$ phase consists of buckled honeycomb structure with half-occupied $\alpha$ and $\beta$ sites as shown in Fig. 1(b) and shows a superior ionic conductivity enabled by the low energy barrier between these sites [20-22].

In our previous study, thin-film growth of the single-phase $AgCrSe_2$ is achieved by careful tuning of growth temperature and composition of supply on the lattice-matched (111) plane of yttria-stabilized zirconia (YSZ) substrate [23]. The 2D atomic arrangement at the interface between $AgCrSe_2$ thin film and YSZ(111) substrate is illustrated in Fig. 1(d). On the YSZ(111) surface, the topmost O atoms form a triangular sublattice with the O-O distance of $d_{\text{O-O}}$ = 3.639 Å. The lattice mismatch with the Se-Se distance ($d_{\text{Se-Se}}$ = 3.680 Å) of $AgCrSe_2$ is $(d_{\text{Se-Se}} - d_{\text{O-O}})/d_{\text{O-O}} = -1.1\%$. This small lattice mismatch enables the $c$-axis-oriented growth of single-phase $AgCrSe_2$ [23]. On the other hand, our previous work suggests the presence of four types of domains: in-plane twisted domains (0° or 60°) and polar domains (+$Z$ polar or −$Z$ polar), indicating difficulty in lifting the nearly degenerate domains. Here, the 0° domain is defined by the in-plane epitaxial relationship of $AgCrSe_2[10\bar{1}0]$ // YSZ$[1\bar{1}0]$ whereas the 60° domain is defined by $AgCrSe_2[10\bar{1}0]$ //

YSZ[$10\bar{1}$], and +$Z$ and −$Z$ polar domains correspond to the $c$ axis of $AgCrSe_2$ pointing upward and downward, respectively. By taking the second layer into account, however, the Zr/Y sublattice makes the (111) plane of YSZ threefold symmetry. Therefore, it is supposed that the thermodynamic stability of the twisted domain nucleation is not perfectly degenerate for the 0° and 60° orientations. Another difficulty relies on the high volatility of Ag compared with Cr. Variation of Ag content in the obtained thin films depends on $T_{sub}$, which hinders the formation of single-phase $AgCrSe_2$ when exploring the thermodynamically stable conditions.

In this study, we have identified the optimal growth temperature and Ag/Cr composition ratio of supply to obtain nearly single domain of $AgCrSe_2$ on YSZ(111) substrate. We have also investigated surface structure and revealed a mixed polar orientation, indicating the possibility of ferroelectricity of $AgCrSe_2$. From the dome-like dependence of volume fraction of twisted domains on the growth temperature, we find the severe growth window in terms of thermodynamics and kinetics in the heteroepitaxy of $AgCrSe_2$ on the YSZ(111) substrate.

## II. EXPERIMENTAL METHODS

For thin-film growth using pulsed-laser deposition (PLD), Ag-rich targets were used to compensate for Ag deficiency with the aim of realizing the high-temperature deposition. The two PLD targets were prepared by mixing $AgCrSe_2$ and $Ag_2Se$ ($P2_12_12_1$) powders with the 2:3 and 2:5 molar ratios, and pressing them into pellets with a diameter of 10 mm. The $AgCrSe_2$ powders were obtained by pulverizing the single-phase polycrystalline $AgCrSe_2$ samples synthesized by solid-state reaction. Details of fabrication and characterization of the polycrystalline $AgCrSe_2$ samples were described in Ref. [23]. The Ag/Cr atomic ratios of the prepared targets consisting of a mixture of $AgCrSe_2$ and $Ag_2Se$ is 4 and 6.

The 100-nm-thick thin films were deposited on YSZ(111) substrates by ablating the targets. Substrate temperature was monitored by an infrared optical pyrometer through viewport. The structural characterization of thin-film samples was performed by x-ray diffraction (XRD, Empyrean, PANalytical B.V.) with a Cu–Kα1 x-ray source ($\lambda$ = 1.5406 Å) and atomic force micrography (AFM, Hitachi High-Tech). The Ag/Cr chemical composition ratio of thin films was determined by energy dispersive x-ray spectroscopy (EDX) using a scanning electron microscope (TM4000Plus II, Hitachi High-Tech) equipped with an energy dispersive spectrometer (AztecOne, Oxford Instruments) and calibrated by inductively coupled plasma optical emission spectroscopy (ICP-OES, Optima 8300, Perkin Elmer). Polar orientation of $AgCrSe_2$ was analyzed by time-of-flight low-energy atom scattering spectroscopy (TOFLAS, TOFLAS-3000, Pascal Co., Ltd.). A neutral He-atom beam accelerated to 3 keV was irradiated on the surface of the $AgCrSe_2$ thin film and the amount of the scattered ions and neutrals was detected as a function of time of flight (TOF) in vacuum chamber with a base pressure below $2\times10^{-3}$ Pa and a sample-to-detector distance of 365 mm. The amount of the scattered ions and neutrals were integrated between TOFs from 3.35 to 3.60 μs to make pole figure, where radial and tangential directions correspond to polar and azimuthal angle of the incident He-atom beam with respect to the normal to the film surface. Crystal structures and two-dimensional atomic arrangements were drawn using VESTA3 [24]. Cubic crystals (such as YSZ and Ag metal) are described using three-index (Miller index) notation, whereas hexagonal crystals (such as $AgCrSe_2$ and $Cr_2Se_3$) is described using four-index (Miller-Bravais index) notation.

## III. RESULTS

### A. Optimization of growth temperature for obtaining stoichiometric composition

We first optimize growth temperature $T_{sub}$ for each Ag-rich target in view of the Ag/Cr composition ratio and surface roughness of the films. Figure 2(a) shows Ag/Cr atomic ratio of the thin films (denoted as "film" Ag/Cr ratio) prepared using targets with different Ag/Cr composition (denoted as "target" Ag/Cr ratio) as a function of $T_{sub}$. As reported previously [23], the film growths using the stoichiometric target and Ag-rich target with target Ag/Cr ratio of 2 result in significant deficiency of Ag (black and blue circles). The stoichiometric film is obtained only when using the target Ag/Cr ratio of 2 and $T_{sub}$ ~ 414°C ($T_{sub}$ was corrected from Ref. [23]). In this study, the use of targets with richer Ag content such as the target Ag/Cr ratio of 4 enables obtaining stoichiometric films at elevated temperatures around 590°C (red circles). Further increase of $T_{sub}$ to be higher than these temperatures, however, results in decrease of film Ag/Cr ratio again due to re-evaporation of Ag. For target Ag/Cr ratio of 6, the stoichiometric film is difficult to obtain because of the sudden drop of film Ag/Cr ratio between $T_{sub}$ of 680°C and 690°C (green circles).

Figure 2(b) shows $T_{sub}$ dependence of the root-mean-square (RMS) value of the surface roughness for the films prepared using target Ag/Cr ratio of 2 (blue), 4 (red) and 6 (green). The RMS for the target Ag/Cr ratio of 4 exhibits the lowest value of about 1.64 nm at $T_{sub}$ ~ 586°C, which corresponds to the temperature where the stoichiometric film Ag/Cr ratio is obtained (orange shaded area). On the other hand, the RMS value for $T_{sub}$ ~ 700°C with target Ag/Cr ratio of 6 is as large as 40 nm. Figure 2(c) shows typical AFM images taken for $T_{sub}$ = 414, 586, and 698°C for target Ag/Cr ratio of 2, 4 and 6, where the film Ag/Cr ratio is nearly stoichiometric. Surface flatness is greatly improved by increasing $T_{sub}$ from 414°C to 586°C with appearance of some triangular-shape step-and-terrace structures. On the other hand, many precipitates emerge on the surface for $T_{sub}$ = 698°C with target Ag/Cr ratio of 6, which can be ascribed to Ag segregation as observed in SEM-EDX mapping (see Supplementary Fig. S1). These results indicate that the 2D growth mode resides in the narrow window of $T_{sub}$ and Ag amount of supply.

Figure 2(d) shows the results of out-of-plane 2theta-omega scan of XRD for the films prepared at the respective $T_{sub}$ of 498 (dark red), 586 (red) and 699°C (orange) with target Ag/Cr ratio of 4, and 698°C (green) with target Ag/Cr ratio of 6. As a reference [23], the data for the previously-reported single-phase $AgCrSe_2$ film for $T_{sub}$ = 414°C with target Ag/Cr ratio of 2 is also plotted (blue). Judging from the observation of the clear ($000\underline{3l}$) diffraction peaks below $T_{sub}$ = 586°C, the *c*-axis-oriented single-phase $AgCrSe_2$ ($R3m$) thin film is obtained. In addition, the peak intensity for $T_{sub}$ = 586°C is slightly larger than that for $T_{sub}$ = 414°C, as a consequence of improvement of the crystallinity by high-temperature growth. Further increase of $T_{sub}$, however, induces decomposition into pure Ag (fcc, $Fm\bar{3}m$) and $Cr_2Se_3$ phases regardless of the value of film Ag/Cr ratio, in addition to development of the $AgCrSe_2$(009) diffraction peak as shown for $T_{sub}$ ~ 700°C for target Ag/Cr ratio = 4 (orange) and 6 (green). These AFM and XRD results indicate that the sharp drop of film Ag/Cr ratio across unity around $T_{sub}$ ~ 685°C observed in Fig.

2(a) is associated with decomposition of $AgCrSe_2$ and re-evaporation of segregated Ag.

**B. Characterization of twisted domain formation by in-plane XRD patterns**

To examine the effect of $T_{sub}$ variation on the twisted domain formation, we first measure in-plane rotation XRD of the $AgCrSe_2$ phase in the thin films for different $T_{sub}$. Here, we exclude decomposition regime $T_{sub}$ > 685°C (black shaded area) from our discussion. Figure 3(a) shows phi-scan XRD patterns of $(10\bar{1}7)$ or $(01\bar{1}\bar{7})$ of $AgCrSe_2$ [denoted as $AgCrSe_2(10\bar{1}7)$ for simplicity, hereafter] for $T_{sub}$ = 414 (blue) with target Ag/Cr = 2, 586°C (red) with target Ag/Cr = 4, and $T_{sub}$ = 698°C (green) with target Ag/Cr = 6, and that (400) of YSZ (the bottom panel). When using target Ag/Cr = 2, the sixfold diffraction peaks are observed for $T_{sub}$ = 414°C despite the threefold $R3m$ symmetry on the $c$-plane of $AgCrSe_2$ (Ref. [23]). In contrast, threefold strong peaks of $AgCrSe_2(10\bar{1}7)$ are observed for $T_{sub}$ = 586°C, which is intrinsic to the $AgCrSe_2$ single crystal. In addition, weak threefold symmetric peaks are still observed at 0°, indicating the presence of the small fraction of 0° domains. By considering that the strong $AgCrSe_2(10\bar{1}7)$ peaks are shifted in phi by 60° with respect to the threefold YSZ(400) peaks, we conclude that epitaxial relationship of the main phase of $AgCrSe_2$ is twisted by 60°, where the $a$ axis of $AgCrSe_2$ is parallel to YSZ$[10\bar{1}]$ and the $c$ axis points upward (+$Z$ polar) or the $a$ axis of $AgCrSe_2$ is parallel to YSZ[$\bar{1}$10] and the $c$ axis points downward ($-Z$ polar) [see Fig. 1(d)]. With further increase of $T_{sub}$ to 652°C (orange), the intensity of the 0° domain develops and becomes half of that of the 60° domain.

To quantitatively evaluate volume fraction of the twisted domains, $T_{sub}$ dependence of the XRD intensity ratio $I_{60°}/(I_{0°} + I_{60°})$ is plotted in Fig. 3(b). Here, $I_{0°}$ and $I_{60°}$ represent integral intensities of the XRD peaks for $AgCrSe_2(10\bar{1}7)$ stemming from the 0° and 60° domains, respectively. Overall, the intensity ratio increases with increasing $T_{sub}$, reaches a maximum of 0.9 at $T_{sub}$ around 590°C, and decreases with further increase of $T_{sub}$. Therefore, we conclude that the 60° domain is most suppressed at $T_{sub}$ ~ 590°C. Such dome-like dependence of $I_{60°}/(I_{0°} + I_{60°})$ on $T_{sub}$ implies severe growth condition and delicate balance between thermodynamic stability and growth kinetics for obtaining the single-domain $AgCrSe_2$ on the YSZ(111) substrate.

**C. Characterization of polarization domain by surface structural analysis**

Next, polar orientation of the optimal thin film is investigated by surface structural analysis using TOFLAS. Figures 4(a) and 4(b) show an illustration of the atomic arrangement of $AgCrSe_2(0001)$ plane ($+Z$ polar surface) viewed from the Ag atomic plane and simulated pole figure of TOFLAS intensity. Figures 4(c) and 4(d) show those for $AgCrSe_2(000\bar{1})$ plane ($-Z$ polar surface). As shown in the simulated TOFLAS images, the difference in polarity of the $AgCrSe_2$ structure can be distinguished by numbers of outer black threefold symmetric spots indicated by white arrows; the (0001) plane exhibits a pair of spots as shown in Fig. 4(b), whereas the $(000\bar{1})$ plane exhibits a single spot as shown in Fig. 4(d). Figure 4(e) shows the experimentally obtained pole figure of the integrated scattering intensity of TOF spectrum for the thin film grown at $T_{sub} = 586$°C with target Ag/Cr ratio = 4. The threefold symmetric image is apparently observed, being consistent with the nearly single domain formation as discussed in Fig. 3. Strong inner spots are observed, which is commonly present in the simulated images of the (0001) and $(000\bar{1})$ planes. In addition, weak three sets of spot-like contrasts can be distinguished as indicated with a dashed circle. We consider that two of them correspond to the black spots from the (0001) plane in Fig. 4(b); one of them corresponds to that from the $(000\bar{1})$ plane in Fig. 4(d). Figure 4(f) shows a sum of the simulated TOFLAS images from the (0001) and $(000\bar{1})$ planes. Some characteristic contrasts including outer spots are consistent with the observation, indicating that the polar domain is not fully aligned. It turns out that there is no preferential poling direction in cooling from $T_{sub}$ across the structural transition temperature $T_s$. Although the ferroic polarization has not been addressed so far for $AgCrSe_2$ in both bulk and thin-film forms, ferroelectric switching was observed in exfoliated flakes of $CuCrSe_2$ [25]. The coexistence of $+Z$ and $-Z$ polar domains in $AgCrSe_2$ thin film indicates the degeneracy of their polar orientations and the possibility of polar switching associated with Ag ionic motion by, for example, external electric field.

## IV. DISCUSSION

Finally, we discuss $T_{sub}$ dependence of twisted and polar domain formation in terms of their energetic perspective qualitatively. Figure 5 shows schematic energy profile along reaction coordinate treating the domain nucleation with its crystal orientation along 0° and 60° with respect to the YSZ(111) substrate as local minima. From the observation that the 60° domain dominates at moderate temperature of $T_{sub} \sim 590$°C, the formation energy of the 60° nuclei can be assumed to be slightly lower than that of the 0° nuclei with energy difference of $\Delta H$. At low $T_{sub}$, both 0° and 60° domains start to grow but the thermal activation $k_B T_{sub}$ of the 0° domain is insufficient to rotate into the most stable orientation. Here, $k_B$ is the Boltzmann constant. At moderate $T_{sub} \sim 590$°C, the thermal energy is large enough to overcome the energy barrier $\Delta E$, promoting formation of the more stable 60° domain, promoting rotation of the metastable 0° domain to the most stable 60° orientation [26]. As a result, the large fraction of the 60° domain develops as observed for $T_{sub} \sim 590$°C For more quantitative analysis, the concrete values of $\Delta H$ and $\Delta E$ should be evaluated by considering the bonding energy between the precursors and substrate surface using, for example, first principles calculations [10]. After the thin-film growth and cooling $T_{sub}$ below the structural transition temperature $T_s$ (indicated by a red arrow), appearance of both $+Z$ and $-Z$ polar domains as observed in Fig. 4, indicate that these two polar orientations are nearly degenerate in energy.

## V. SUMMARY

In summary, we have investigated thermodynamic stability of the twisted domain formation in the epitaxial thin-film growth of $AgCrSe_2$ in pulsed-laser deposition, in combination of variation of growth temperature and compensation for Ag deficiency. By increasing the target Ag/Cr ratio from 2 to 4, the optimal $T_{sub}$ for obtaining the stoichiometric film is increased from 414 to 586°C. The out-of-plane XRD analysis reveals that $c$-axis oriented growth of single-phase $AgCrSe_2$ thin films is obtained on the lattice-matched YSZ(111) substrate for $T_{sub}$ = 414 and 586°C while the decomposition into Ag and $Cr_2Se_3$ occurs for $T_{sub}$ around 685°C. The in-plane XRD shows that only the moderate temperature of $T_{sub}$ = 586°C yields almost single-domain $AgCrSe_2$ with in-plane epitaxial relationship of $AgCrSe_2[10\bar{1}0]$ // YSZ$[10\bar{1}]$. Based on these observations, we consider an energy landscape for the metastable 0° domain and the most stable 60° domain. In addition, surface analysis using TOFLAS measurement shows that the $AgCrSe_2$ film consists of both $+Z$ and $-Z$ polar orientations, indicating the possible ferroelectric properties owing to the Ag sites. Our findings bring valuable insight to thermodynamics and kinetics of thin-film growth for various two-dimensional compounds possessing rhombohedral crystal structures.

## ACKNOWLEDGMENTS

The ICP-OES analysis was carried out by Shigeru Tamiya at Core Facility Center, the University of Osaka as a part of the MEXT Project for promoting public utilization of advanced research infrastructure (a program for supporting the construction of core facilities), Grant No. JPMXS0441200023. The simulated TOFLAS images were drawn by Pascal Co., Ltd. This work was supported by JST, PRESTO Grant No. JPMJPR21A8 and JSPS KAKENHI, Grant Nos. JP23H01686, JP23K26379, JP23K22453, JP24K21531, Murata Science and Education Foundation, Iketani Science and Technology Foundation, and The Thermal & Electric Energy Technology Foundation.

## ORCID iDs

Kenshin Inamura – https://orcid.org/0009-0004-3536-8398

Kazutaka Kudo – https://orcid.org/0000-0003-2839-6791

Jobu Matsuno – https://orcid.org/0000-0002-3018-8548

Junichi Shiogai – https://orcid.org/0000-0002-5464-9839

## AUTHOR DECLARATIONS

### Conflict of Interest

The authors have no conflicts to disclose.

**Data availability**

The data that support the findings of this study are available from the corresponding author upon reasonable request.

**Figure caption**

**FIG. 1.** (a)(b) Schematic crystal structures of $AgCrSe_2$ in (a) low temperature and polar phase in $R3m$ (left) structure and (b) high temperature and nonpolar phase in $R\bar{3}m$ structure. The red, blue, and green spheres represent Ag, Cr, and Se, respectively. The red and white bicolor spheres in (b) represent the half occupancy at the Ag sites in $R\bar{3}m$ structure. (c) Cross-sectional view of the Ag atomic layer intercalated between $CrSe_2$ networks for +*Z*-polar (top panel) and −*Z*-polar (bottom) orientation. (d) Top-view schematic of epitaxial relationship at interface between the *c*-plane of $AgCrSe_2$ and the (111) plane of the YSZ substrate. Green shaded area represents the $AgCrSe_2$ unit cell with the epitaxial relationship of either $a$ or $b$ axis of $AgCrSe_2$ being parallel to YSZ$[1\bar{1}0]$ (denoted as 0° domain) depending on the polar orientation. Red shaded area represents the epitaxial relationship of either $a$ or $b$ axis of $AgCrSe_2$ being parallel to YSZ$[10\bar{1}]$ (denoted as 60° domain) depending on the polar orientation.

**FIG. 2.** (a)(b) Growth temperature $T_{sub}$ dependence of (a) Ag/Cr composition ratio and (b) the root-mean-square (RMS) value of surface roughness of thin-film samples prepared from the target Ag/Cr ratio of 1 (black), 2 (blue), 4 (red), and 6 (green). (c) Surface morphology grown at $T_{sub}$ = 414°C (left panel), 586°C (center), and 698°C (right) using

target Ag/Cr ratio of 2, 4, and 6, respectively. The numbers in the images represent the RMS value of surface roughness. (d) Out-of-plane x-ray diffraction pattern for $T_{sub}$ = 414°C (with target Ag/Cr ratio of 2, blue), 498, 586, and 699°C (with target Ag/Cr = 4, red), and 698°C (with target Ag/Cr = 6, green).

**FIG. 3.** (a) In-plane phi-scan of x-ray diffraction (XRD) patterns for $AgCrSe_2(10\bar{1}7)$ with $T_{sub}$ = 414°C with target Ag/Cr ratio of 2 (blue), and 586°C (red) and 652°C (orange) with target Ag/Cr ratio of 4, and that for YSZ(400). (b) The growth temperature $T_{sub}$ dependence of $I_{60°}/(I_{0°} + I_{60°})$ with $I_{0°}$ and $I_{60°}$ being the integral intensities of the XRD peaks of the 0° and 60° domains.

**FIG. 4.** (a) In-plane of atomic arrangement and (b) simulated TOFLAS pole figure for the (0001) plane of $AgCrSe_2$. (c)(d) Those for the $(000\bar{1})$ plane of $AgCrSe_2$. (e) Experimentally obtained TOFLAS image for $AgCrSe_2$ thin film grown at $T_{sub}$ = 586°C with target Ag/Cr ratio of 4. (f) The simulated TOFLAS pole figure for a mixture of the (0001) and $(000\bar{1})$ planes of $AgCrSe_2$(0001) plane of $AgCrSe_2$.

**FIG. 5.** Temperature variation of the energy landscape for metastable 0° and most stable

60° orientation of the $AgCrSe_2$ nuclei.

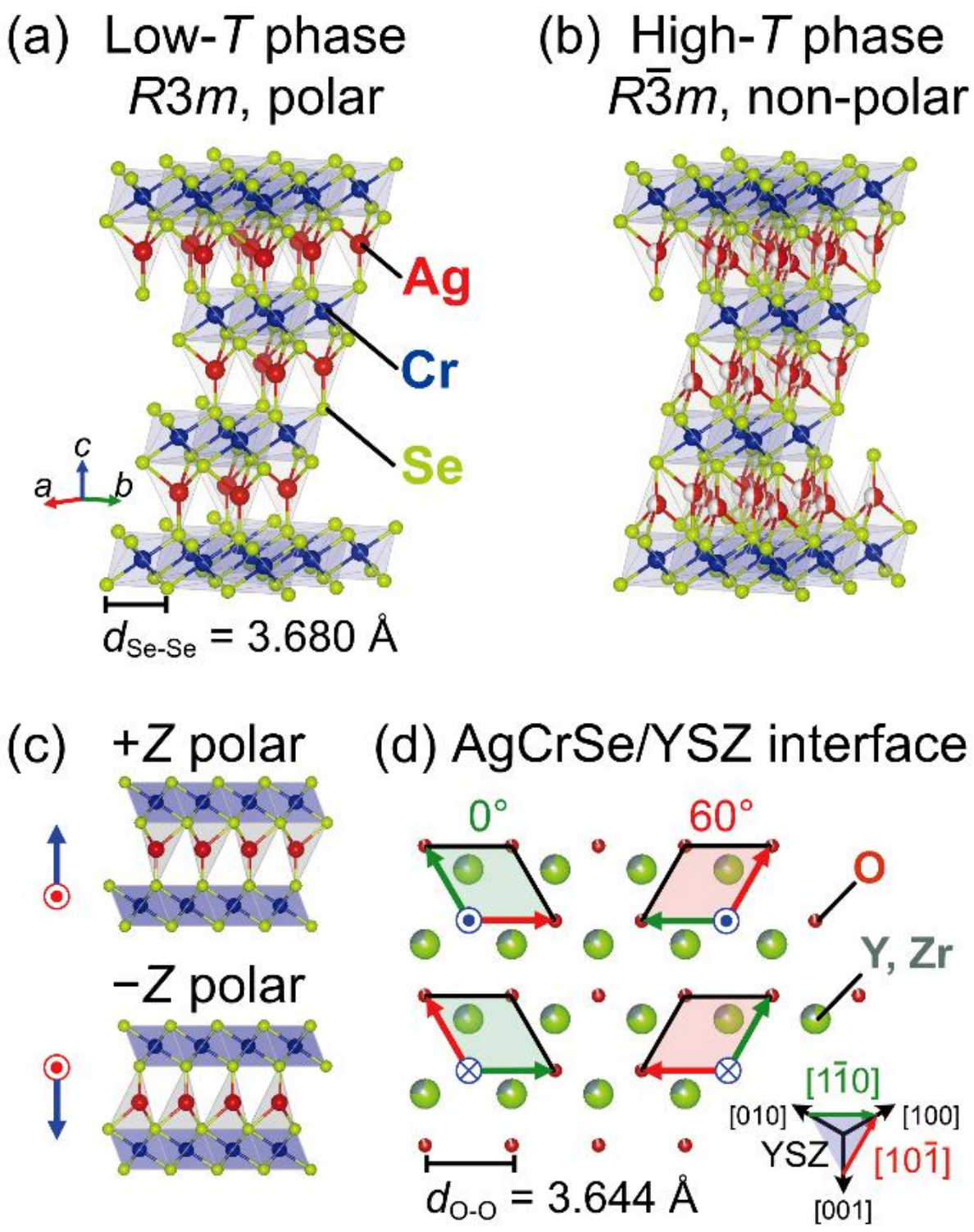

FIG. 1. (single column) Sato et al.

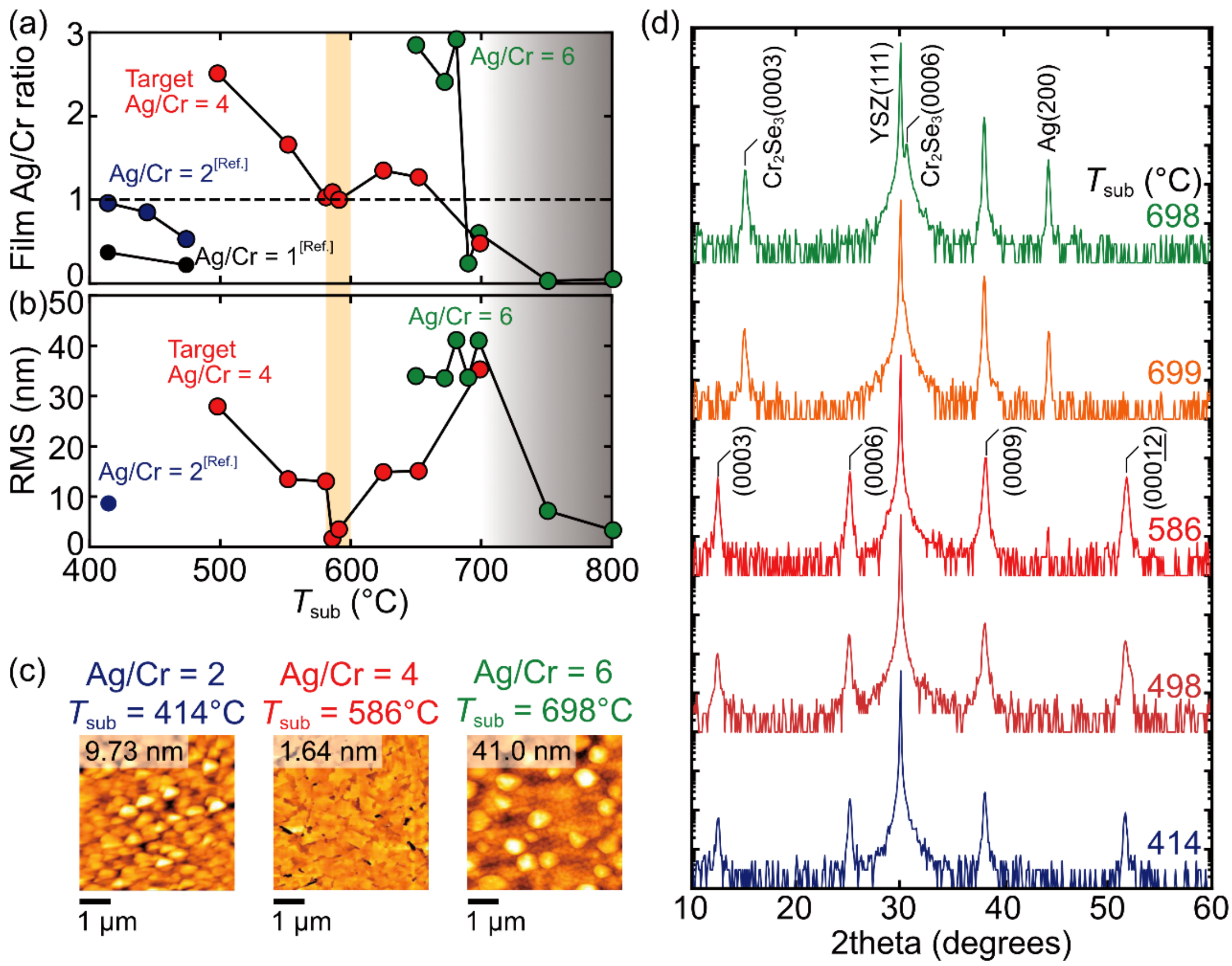

FIG. 2. (double column) Sato et al.

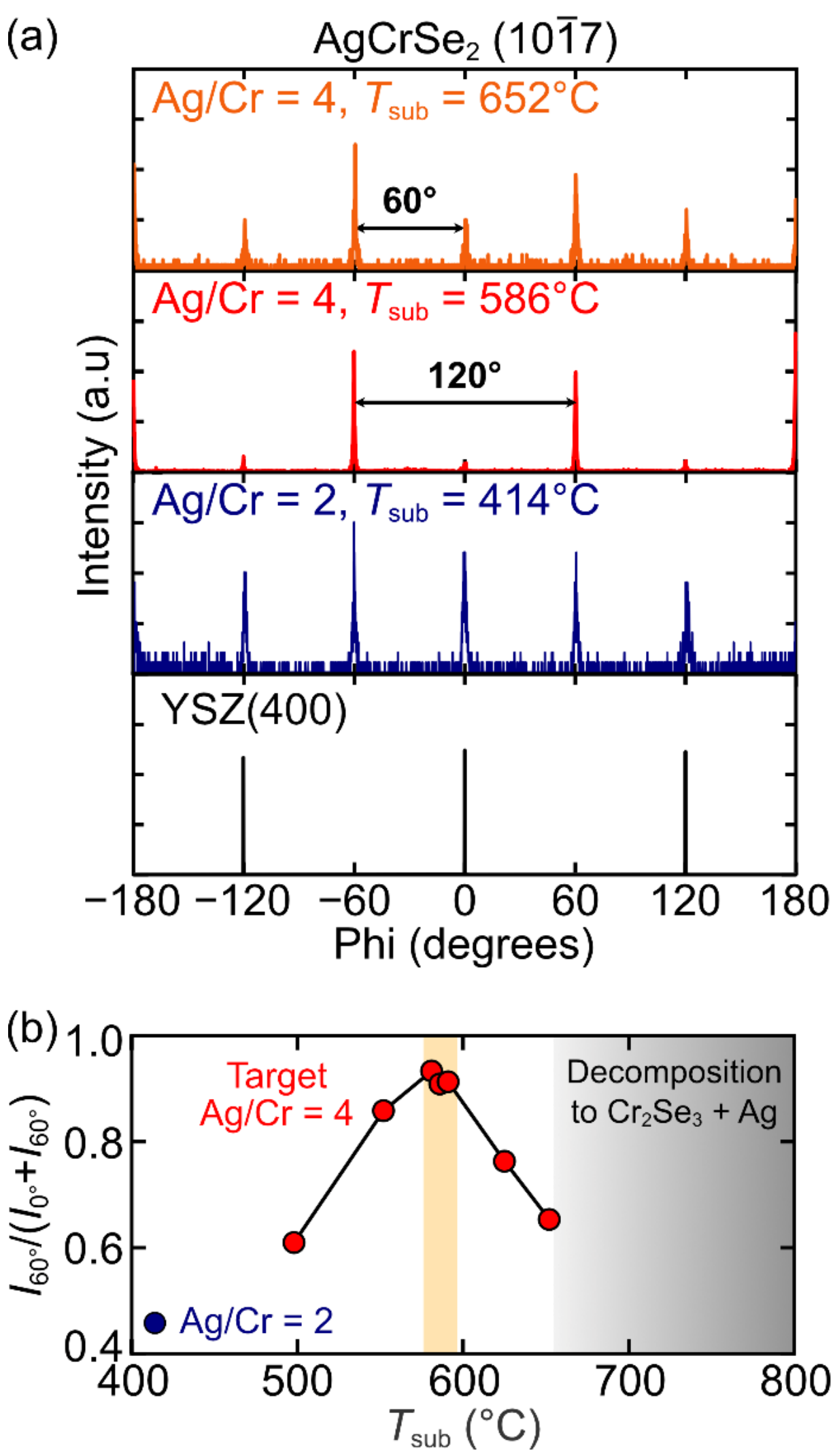

FIG. 3. (single column) Sato et al.

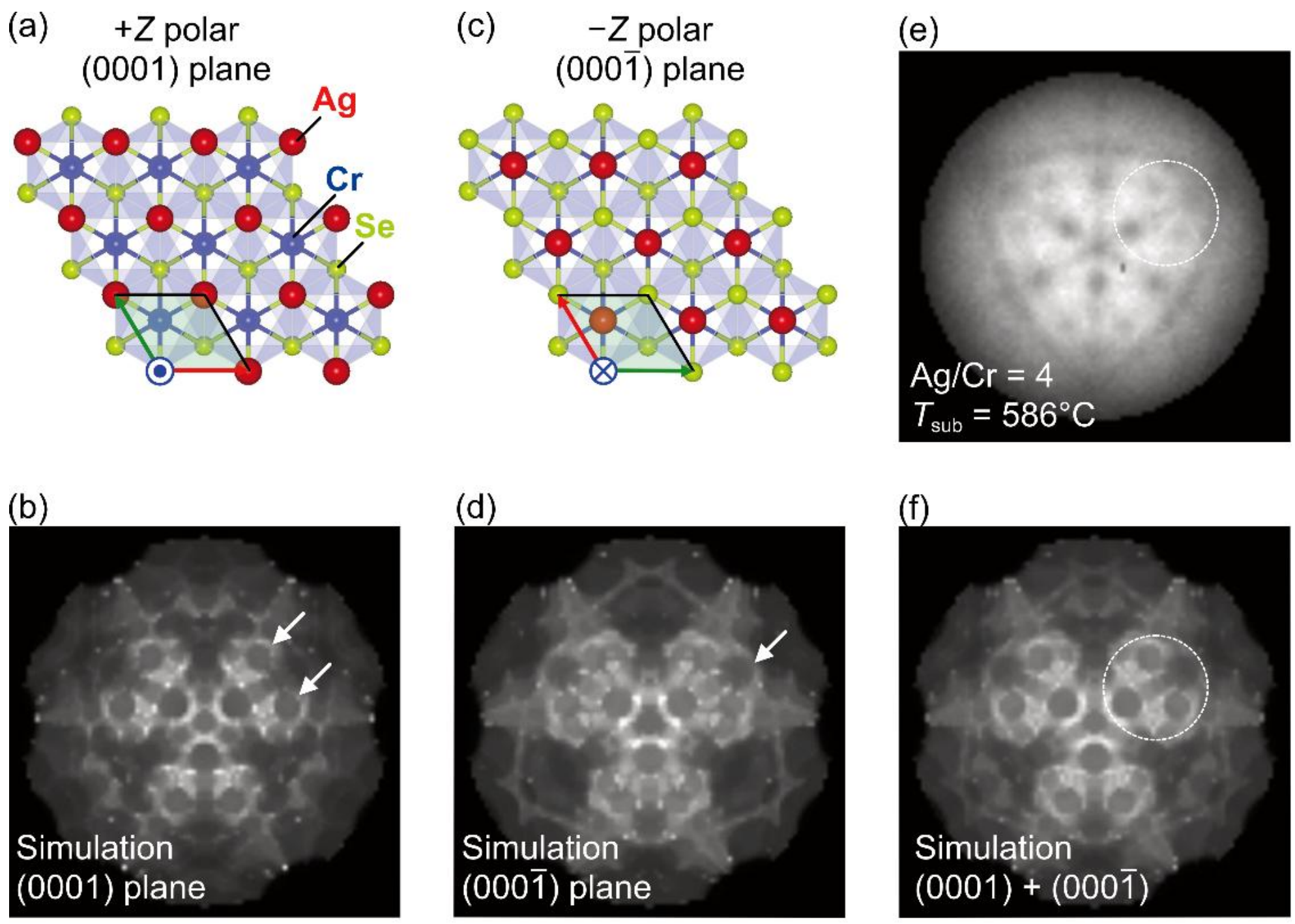

FIG. 4. (double column) Sato et al.

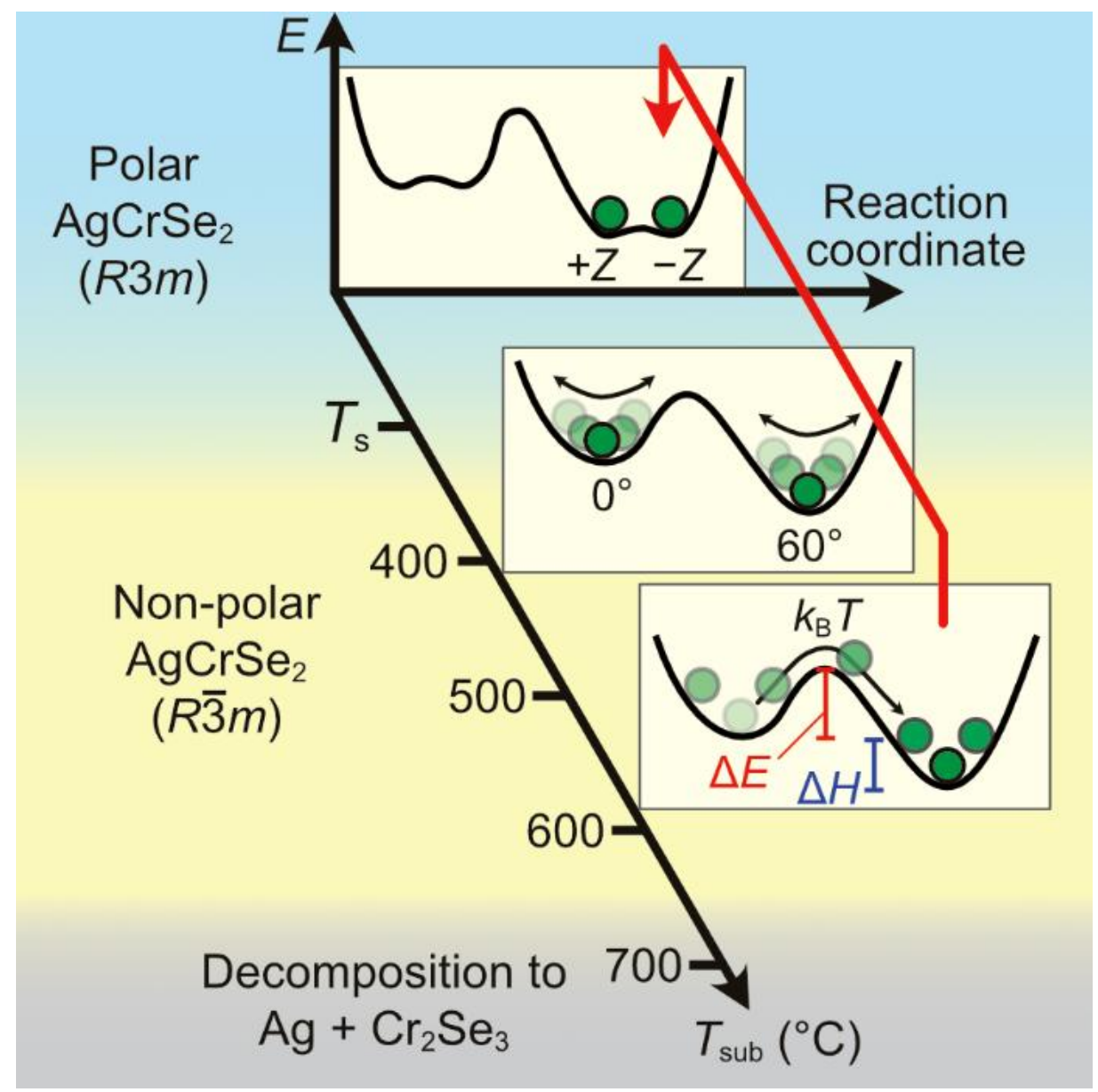

FIG. 5. (single column) Sato et al.

Supplementary Materials for

# Thermodynamic stability of twisted domains in $AgCrSe_2$ thin films grown on lattice-matched YSZ(111) substrate

Haruto Sato[1], Kota Mihara[1], Kenshin Inamura[1], Yusuke Tajima[1],

Kazutaka Kudo[1,2], Jobu Matsuno[1,2], and Junichi Shiogai[1,2,†]

[1]*Depertment of Physics, The University of Osaka, Toyonaka, Osaka, 560-0043, Japan.*

[2]*Division of Spintronics Research Network, Institute for Open and Transdisciplinary Research Initiatives, The University of Osaka, Suita, Osaka, 565-0871, Japan.*

[†]The author to whom correspondence should be addressed.

junichi.shiogai.sci@osaka-u.ac.jp

## 1. Composition mapping of the surface structure of the sample obtained by high-temperature growth

Figure S1 shows scanning electron microscopy (SEM) image and composition mapping obtained by energy dispersive x-ray spectroscopy (EDX) for the film grown at $T_{sub}$ = 698°C with target Ag/Cr ratio of 6. For the sake of avoiding the effect of substrate, we employed $Al_2O_3$ substrate for composition analysis. As can be seen, the precipitates seen in the SEM image can be assigned to the Ag segregations.

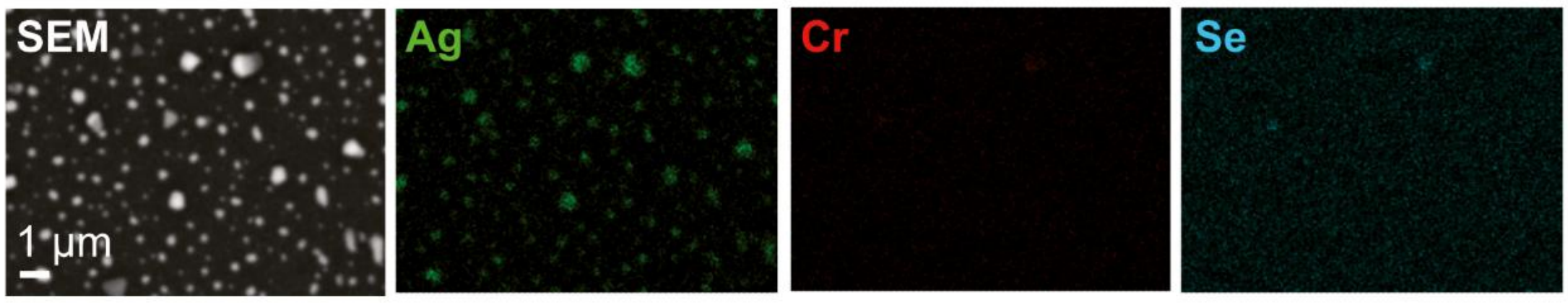

FIG. S1. SEM image (Left panel) and EDX composition mapping of Ag (green), Cr (red), and Se (blue) for the film grown at $T_{sub}$ = 698°C with target Ag/Cr ratio of 6 on $Al_2O_3$ substrate.